\documentclass[pdftex,twocolumn,epjc3,a4paper]{svjour3}
\RequirePackage[T1]{fontenc}
\smartqed  
\usepackage{amsmath}
\usepackage{amssymb}
\RequirePackage{flushend}
\RequirePackage{mathrsfs}
\RequirePackage{dsfont}
\usepackage{units}
\usepackage{graphicx}
\usepackage[colorlinks = true,
            linkcolor = blue,
            urlcolor  = blue,
            citecolor = blue,
            anchorcolor = blue]{hyperref}
\usepackage{cite}
\usepackage{xcolor} 
\usepackage{upgreek}
\usepackage[switch]{lineno}
\journalname{Eur. Phys. J. C}
\newcommand{\mass}{0.175\,g\,}

\begin{document}

\author
{%
A.~H.~Abdelhameed\thanksref{addr1} \and
G.~Angloher\thanksref{addr1}\and
A.~Bento\thanksref{addr1,addr2}\and
E.~Bertoldo\thanksref{addr1,addr3}\and
A.~Bertolini\thanksref{addr1}\and
L.~Canonica\thanksref{t1,addr1}\and
N.~Ferreiro~Iachellini\thanksref{addr1}\and
D.~Fuchs\thanksref{addr1}\and
A.~Garai\thanksref{addr1}\and
D.~Hauff\thanksref{addr1}\and
A.~Nilima\thanksref{addr1}\and
M.~Mancuso\thanksref{addr1}\and
F.~Petricca\thanksref{addr1}\and
F.~Pr\"obst\thanksref{addr1}\and
F.~Pucci\thanksref{addr1}\and
J.~Rothe\thanksref{addr1,addr4}\\
}

\institute
{%
Max-Planck-Institut f\"ur Physik, D-80805 M\"unchen, Germany 
\label{addr1}  
\and
also at: LIBPhys-UC, Departamento de Fisica, Universidade de Coimbra, P3004 516 Coimbra, Portugal 
\label{addr2}
\and
now at: Institut de F\'{i}sica d'Altes Energies (IFAE), Barcelona Institute of Science and Technology (BIST), Bellaterra (Barcelona) E-08193, Spain
\label{addr3}
\and
now at: Physik-Department and Excellence Cluster Universe, Technische Universit\"at M\"unchen, D-85748 Garching, Germany 
\label{addr4}
}

\thankstext[$\star$]{t1}{Corresponding author: canonica@mpp.mpg.de}

\title{A low-threshold diamond cryogenic detector for sub-GeV Dark Matter searches}

\date{Received: date / Accepted: date}

\maketitle

\begin{abstract}
In this work we report the realization of the first low-threshold cryogenic detector that uses diamond as absorber for astroparticle physics applications. We tested two 0.175$\,$gr CVD diamond samples, each instrumented with a W-TES. The sensors showed transitions at about 25 mK. 
We present the performance of the diamond detectors and we highlight the best performing one, where we obtained an energy threshold as low as  16.8 eV. 
This promising result lays the foundation for the use of diamond for different fields of applications where low threshold and excellent energy resolution are required, as i.e. light dark matter searches and BSM physics with coherent elastic neutrino nucleus scattering.


\end{abstract}
Compiled on \today

\section{Introduction}
It is undeniable that the identification of dark matter (DM) is one of the most urgent open questions in physics today. 
Several observational evidences based on gravitational effects (i.e. anisotropies in the cosmic microwave background, large-scale structure distributions, galaxy cluster velocity dispersions, gravitational lensing effect) support the assumption that ~26\% of the mass-energy density in the Universe is in the form of non-barionic and cold DM.  For recent reviews see, e.g., \cite{Massey_2010, Salucci2018}.  However, despite the world-wide efforts and the tremendous experimental progresses, the nature of DM is still unresolved. 
A standard approach in the field of direct DM searches is to look for the scattering of hypothetical DM particles with the atomic nuclei of target materials, contained in low-background experiments typically located in underground laboratories. One of the crucial parameters in direct DM experiments is the energy threshold, since it drives the sensitivity to the detection of low-energy nuclear recoils induced by DM interactions. 
While the elastic scattering of DM particles with masses above a few GeVs can be easily accessed by experiments using large TPCs filled with noble-liquids (Argon \cite{ DarkSide50, DEAP3600_PhysRevD.100.022004} and Xenon\cite{LUX2017, PandaX-II2017, Xenon1ton2018}), light DM particles with masses in the sub-GeV range are not accessible to these experiments due to the kinematics of the process. Simply, light DM particles, that interact with Ar or Xe nuclei, induce nuclear recoils that are below the experimental threshold. Recently, the Migdal effect \cite{PhysRevLett.121.101801, Ibe2018} has been proposed to extend the reach of these experiments to sub-GeV DM masses \cite{XENON_Migdal2019, CDEX_Migdal2019}, but the existence of this effect has still to be experimentally demonstrated. 
An experimental technique complementary to TPCs is the one of cryogenic detectors. They enable the reach of low-energy thresholds by means of detecting of the phonons created following the particle interaction with the crystalline detector nuclei. As of today, Si\cite{SuperCDMS_Silicon}, Ge \cite{EDELWEISS_2019} and CaWO$_4$ \cite{CRESST2019} crystals are the targets most widely used in state-of-the art experiments for direct DM searches. 
For the read-out of the tiny temperature variations induced by particle interactions, the crystal targets must be instrumented with highly sensitive temperature sensors. Among the available thermometers, Transition Edge Sensors using W films\cite{TESCabrera} and Neutron Transmutation Doped thermistors\cite{PirroNTD} have the properties that fulfill the requirements of a sub-GeV DM experiment, having obtained energy thresholds of  \textit{O}(10 eV). Thanks to their performance, cryogenic detector are currently leading the sensitivity of spin-independent DM-nucleons interactions for DM masses below few GeV \cite{SuperCDMS_Silicon, EDELWEISS_2019, CRESST2019} .
However, to further extend the reach of cryogenic experiments and explore new region of parameters of lighter DM particles, experimental thresholds as low as few eV are required. \\
Diamond targets have the potential to reach such a low energy threshold thanks to their superior thermal properties \cite{PhysRevLett.16.354} and only recently it has been proposed as detection medium for sub-GeV DM searches \cite{PhysRevD.99.123005}.
Despite the outstanding thermal properties, diamond has not been operated as a cryogenic absorber for astroparticle physics applications until recently. This was mainly due to the limited availability of cheap, high-purity and large-size single crystal (SC) diamonds on the market. Our group has already reported the first operation of a diamond cryogenic absorber equipped with a W-TES, but due to the poor performance of the setup, it was not possible to obtain any information on the threshold achieved by the prototype\cite{DiamondLTD18}.
Here we report the results on the cryogenic performance of a new prototype realized with a synthetic SC diamond equipped with a W-TES. We achieved a detector energy threshold of 16.8 eV. This result opens a wealth of opportunities for using high-purity lab-grown diamonds in field of light DM particle searches, but also for other sectors where low energy thresholds are needed (i.e. coherent elastic neutrino nucleus scattering). The low-threshold achieved will allow to study new properties of DM with the data presented in this work. We leave this topic for a future publication that is currently in preparation.

\section{Diamond as cryogenic target for DM detection}
\label{sec:2}
Cryogenic calorimeters have been developed to explore different sectors of particle physics and astrophysics, from direct DM searches \cite{SuperCDMS_Silicon, EDELWEISS_2019, CRESST2019}, to neutrinoless double beta decay  \cite{CUORE_PRL, PRL-Cupid0, AMoRE} and neutrino mass measurements \cite{Echo, Holmes}. One of their characteristics is that they can be realized with many different materials, with the only requirement to have crystalline structure with low enough heat capacity at the operating temperature \textit{O}(10~mK). 

The Debye temperature of diamond, that can be used as a figure of merit of the thermal properties of the material\cite{Kittel1974}, is among the highest available in nature (2220 K) and it is significantly higher than other established target materials used in DM direct detection experiments (e.g. 645 K for Si, 371 K for Ge and 250 K for CaWO$_4$). 
This parameter ensures high phonon density of states, that is key for achieving the ultimate sensitivity in the detection of small energy deposits. Moreover, diamond is made of light nuclei (Z=6) and this ensures a good match in the kinematics of the interaction of the light DM particle. A complete review of the properties of diamond as cryogenic sub-GeV DM detector can be found in~\cite{PhysRevD.99.123005}.\\
Thanks to the developments in the synthetic grown diamond technologies of the last decades, the barrier of the availability of high-quality crystals has been overcome. We can nowadays easily find on the market SC crystals grown using the Chemical Vapour Deposition (CVD) technology\cite{20101}. In this technique the crystal is grown starting from a source gas (typically methane) that is deposited over a proper seed substrate. The samples used in this work have been realized using the heteroepitaxial technique on a foreign multilayer substrate made of Ir/YSZ/Si\cite{AudiaTEc_1}. This, in contrast to the homoepitaxial technique, does not require the use of a diamond seed to start the growth process. However, it requires only a lattice-matched substrate, enabling the growth of large volume diamonds more easily. 

\section{Experimental setup}
\begin{figure}
\centering 
\includegraphics[width=.30\textwidth]{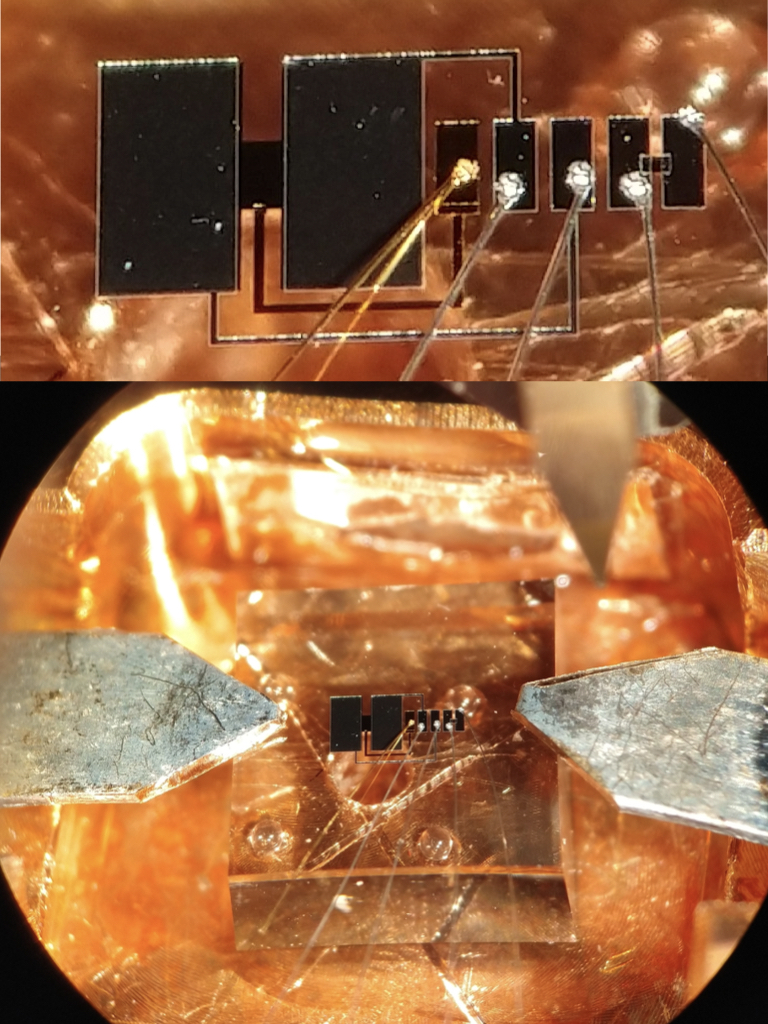}
\caption{
\textbf{Top}: Picture of the TES sensor fabricated on one of the two diamond crystals.  On the left, it is possible to see two large aluminum areas (0.5$\times$1)~mm$^2$ 
evaporated on top of a darker strip of tungsten (140$\times$300) ${\mu} m^2$. 
A 50~nm  thick and 1.5~mm  long strip of gold connects the tungsten layer to a 40~${\mu}$m thick gold pad, to which we bond a 25~$\mu$m gold wire; this gold wire provides a weak thermal coupling to the heat bath at $\sim$10~mK. On the right part of the picture, there is the heater: this is made of a 50~nm thick strip of gold with two aluminum pads deposited on top. 
\textbf{Bottom}: One of the two diamond crystals instrumented with the TES and housed in the copper structure. 
}
\label{fig:foto profilo}
\end{figure}

In this work, the crystals used as absorbers are made by two identical SC diamond samples of \mass each and ($2\times5\times5$)~mm$^3$, grown at AuDiaTec~\cite{audiatec}.
To detect the phonons produced in the absorber following a particle interaction, both crystals are instrumented with a W-TES. The TES design is very similar to the ones used for the CRESST-III experiment~\cite{Rothe2018}. They are made of a thin strip of W with two large Al pads partially overlapping the W layer. These Al pads have two different features: they serve as phonon collectors \cite{TESCabrera} and as ohmic contacts. They are connected via a pair of 25~$\mu$m Al bond wires through which the bias current is injected. The W film is also connected by a long and thin strip of Au to a thicker Au bond pad on which a 25~$\mu$m Au wire is bonded. This connection serves as weak thermal link between the sensor and the heat bath at $\sim$10~mK. On the same surface, but separated from the TES, we also evaporated a heater. The heater is made of a thin strip of Au with two Al pads deposited on top. These pads are also bonded with a pair of 25~$\mu$m Al bond wires through which a tunable current can be injected to maintain the TES at the desired operating temperature. 
The heater is also used to inject artificial pulses to monitor the detector response over time and to refine the energy calibration during data analysis. In Figure~\ref{fig:foto profilo} (top) we show one of the two diamond crystals after the TES fabrication.
\\
Each of the two diamond crystals is housed in a Cu structure, held in position by a pair of bronze clamps (see Figure~\ref{fig:foto profilo} (bottom)). Four sapphire balls sitting on the Cu holder are used as spacers and provide a point-like contact between the crystal and the holder itself. At a distance of about 0.5 cm from the crystal, a $^{55}$Fe  source (activity $\sim$0.3 Bq) is installed for calibration purposes.
The samples were cooled down in a dilution refrigerator at the Max Planck institute for Physics in Munich, Germany, in a surface building, using a Kelvinox-400HA dilution refrigerator from Oxford Instruments (see \cite{MeVscale} and references therein for details of the cryogenic infrastructure).
Electrical connections to the TES were made by superconducting NbTi twisted pairs wires. The TES readout circuit uses a reference resistor at mK and a commercial SQUID system sensor as front end amplifier, combined with a CRESST-like detector control system~\cite{ANGLOHER2009}. The two detectors were readout using an Applied Physics System model 581 DC SQUID.\\

\section{Measurement and data processing}
\label{sec:3}
\begin{figure}[t]
    \centering
    \includegraphics[width=.40\textwidth]{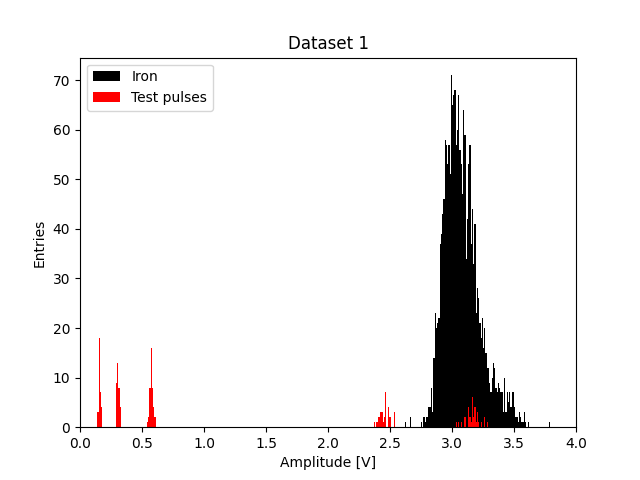}
    \caption{Spectrum of events coming from the iron source (black) and events coming from the artificial pulses sent by a heater (red) in data set 1.}
    \label{fig:ErnieCal}
\end{figure}

\begin{figure}[!ht]
    \centering
    \includegraphics[width=.40\textwidth]{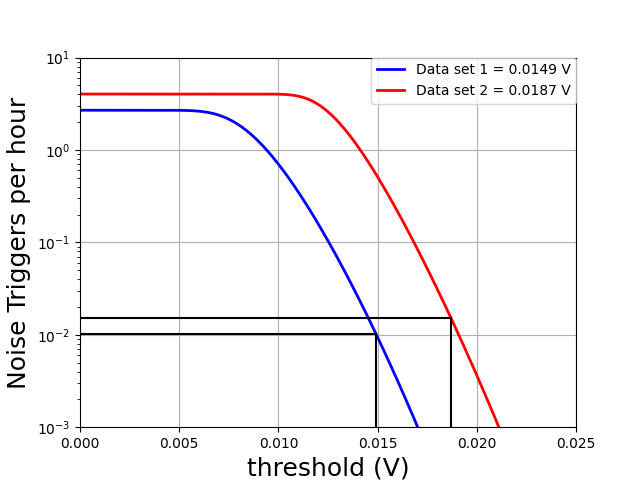}
    \caption{Threshold evaluation of the two sets of data. The threshold is estimated with a function that gives the number of noise triggers for a given threshold (see \cite{Mancuso:2018zoh} for details). The thresholds that have been calculated are 14.9 mV for data set 1 and 18.7 mV for data set 2. Taking into account the calibration functions, they correspond to 19.7 eV and 16.8 eV respectively.}
    \label{fig:3thresh}
\end{figure}

During the cryogenic measurement, the diamond samples were measured simultaneously in a stable working point for 58.4 hours. We collected statistics for a total exposure of $4.3 \times 10^{-4}$ ~kg$\cdot$d. The TESs showed a superconducting transition at about 25 mK. 
In the following, the data sets acquired with the diamond samples \#1 and \#2 are referred to as data set 1 and 2 respectively. \\
Given the small size of the samples and the limited dynamic range of the TESs, a sub-set of measurements was carried out in under-performing conditions. The goal was to extend the sensor dynamic range such that events from the $^{55}$Fe at 5.9~keV were clearly visible and could be used for calibration purposes. The artificial pulses sent through the heater during this measurement were used 
to calculate the conversion factor between the deposited energy and the injected heater voltage in the entire energy range from threshold up to the iron energy region. Thanks to this calibration of the heater response, we were able to reconstruct the energy scale also in different operational conditions, namely during the full data set acquired with the optimized settings. 
As an example, events from the $^{55}$Fe source and from the artificial pulses injected through the heater are shown in Fig.~\ref{fig:ErnieCal} for detector 1.
After the data set for calibration was acquired, the detectors were operated at their optimal operating point, focusing on achieving a low energy threshold.
Data were acquired in a continuously recorded stream and processed with a software trigger after data taking\cite{MeVscale}. \\
Triggered events were filtered with an \textit{optimum filter} \cite{OptimumFilter_GM} algorithm which improves the reconstruction of the amplitude of a pulse with a given pulse shape in the presence of noise.
The noise conditions of each detector were inferred from empty traces which were acquired with the continuous recording process.
For the representation of the pulse shape, 
we created a standard event (SEV) by averaging a selection of clean particle events. Following the directions described in \cite{probst1995model}, the SEV could be used to reproduce the shape of the acquired pulses at different energies. \\
With the aim of estimating the noise level of the baselines of each detector, we superimposed the SEV with a fixed amplitude on a set of empty traces. The distribution of this amplitude evaluated with the filter was then fitted with a gaussian function. The sigma of this distribution was used to determine the resolution of the baseline. The obtained values in Volt and their respective calibrated values in eV can be found in Tab.~\ref{tab:table} for the two diamond samples.\\
Following the approach described in \cite{Mancuso:2018zoh}, the threshold was computed accepting a fixed number of noise triggers for exposure units which is negligible compared to the particle event rate.
For this analysis in both modules the number of accepted noise trigger events 
is around 1\% of the total number of triggered events. 
The final threshold values obtained with this method can be inferred from Fig. \ref{fig:3thresh}, where the noise trigger 
curves are shown as a function of the trigger threshold. Taking into account the calibration factors, these threshold values correspond to 19.7~eV and 16.8~eV for data set 1 and data set 2, respectively.

This result, although obtained with a prototype, shows for the first time the excellent performance of diamond when operated as cryogenic detector. We believe that future work on the optimization of the W-TES design to better match the properties of thermal signals in diamonds will allow for further reduction of the energy threshold, making the goal of reaching \textit{O}(eV) values within reach.

\begin{table}[!ht]
\centering
\begin{tabular}{c|c|c}
\hline
     \textbf{data set} & \textbf{1} & \textbf{2} \\ \hline
    baseline resolution [mV] & 2.68 & 3.81\\ \hline 
    baseline resolution [eV] & 3.54 & 3.42\\ \hline
     threshold [mV] & 14.9 & 18.7\\ \hline
     threshold [eV] & 19.7 & 16.8\\ \hline

\end{tabular}
\caption{\label{tab:table} Values for threshold and baseline resolution of the two data sets. Each value is shown in mV and in eV.}
\end{table}

\section{Conclusions}
This work shows the advanced performance achieved with two detectors made of diamonds operated as cryogenic detectors. The results presented in this work represent a major milestone for future sub-GeV DM investigations, since the low-energy thresholds achieved with these prototypes already compete with the most advanced detector currently operating in the field of light DM searches. A dedicated future work on the exploration of new region of the parameter space for DM interactions with regular matter obtained with the data presented in this work is currently under preparation. 

\begin{acknowledgements}
This research was supported by the Excellence Cluster ORIGINS which is funded by the Deutsche Forschungsgemeinschaft (DFG, German Research Foundation) under Germany's Excellence Strategy – EXC-2094 – 390783311.

\end{acknowledgements}

\bibliographystyle{h-physrev}
\bibliography{biblio.bib}

\begin{thebibliography}{10}

\bibitem{Massey_2010}
R.~Massey, T.~Kitching, and J.~Richard,
\newblock Reports on Progress in Physics {\bf 73}, 086901 (2010).

\bibitem{Salucci2018}
P.~Salucci,
\newblock Foundations of Physics {\bf 48}, 1517 (2018).

\bibitem{DarkSide50}
\textbf{DarkSide} Collaboration, P.~Agnes {\em et~al.},
\newblock Physical Review D {\bf 98}, 102006 (2018).

\bibitem{DEAP3600_PhysRevD.100.022004}
\textbf{DEAP-3600} Collaboration, R.~Ajaj {\em et~al.},
\newblock Phys. Rev. D {\bf 100}, 022004 (2019).

\bibitem{LUX2017}
\textbf{LUX} Collaboration, D.~S. Akerib {\em et~al.},
\newblock Physical Review Letters {\bf 118}, 021303 (2017).

\bibitem{PandaX-II2017}
\textbf{PandaX-II} Collaboration, X.~Cui {\em et~al.},
\newblock Physical Review Letters {\bf 119}, 181302 (2017).

\bibitem{Xenon1ton2018}
\textbf{XENON} Collaboration, E.~Aprile {\em et~al.},
\newblock Physical Review Letters {\bf 121}, 111302 (2018).

\bibitem{PhysRevLett.121.101801}
M.~J. Dolan {\em et~al.},
\newblock Phys. Rev. Lett. {\bf 121}, 101801 (2018).

\bibitem{Ibe2018}
M.~Ibe {\em et~al.},
\newblock Journal of High Energy Physics {\bf 2018}, 194 (2018).

\bibitem{XENON_Migdal2019}
\textbf{XENON} Collaboration, E.~Aprile {\em et~al.},
\newblock Physical Review Letters {\bf 123}, 241803 (2019).

\bibitem{CDEX_Migdal2019}
\textbf{CDEX} Collaboration, Z.~Z. Liu {\em et~al.},
\newblock Physical Review Letters {\bf 123}, 161301 (2019).

\bibitem{SuperCDMS_Silicon}
\textbf{SuperCDMS} Collaboration, I.~Alkhatib {\em et~al.},
\newblock Physical Review Letters {\bf 127}, 061801 (2021).

\bibitem{EDELWEISS_2019}
\textbf{EDELWEISS} Collaboration, E.~Armengaud {\em et~al.},
\newblock Physical Review D {\bf 99}, 082003 (2019).

\bibitem{CRESST2019}
\textbf{CRESST} Collaboration, A.~H. Abdelhameed {\em et~al.},
\newblock Physical Review D {\bf 100}, 102002 (2019).

\bibitem{TESCabrera}
K.~D. Irwin {\em et~al.},
\newblock Review of Scientific Instr. {\bf 66}, 5322 (1995).

\bibitem{PirroNTD}
S.~Pirro and P.~Mauskopf,
\newblock Annual Review of Nuclear and Particle Science {\bf 67}, 161 (2017).

\bibitem{PhysRevLett.16.354}
W.~Saslow {\em et~al.},
\newblock Phys. Rev. Lett. {\bf 16}, 354 (1966).

\bibitem{PhysRevD.99.123005}
N.~Kurinsky {\em et~al.},
\newblock Phys. Rev. D {\bf 99}, 123005 (2019).

\bibitem{DiamondLTD18}
L.~Canonica {\em et~al.},
\newblock Journal of Low Temperature Physics {\bf 199}, 606 (2020).

\bibitem{CUORE_PRL}
\textbf{CUORE} Collaboration, C.~Alduino {\em et~al.},
\newblock Phys. Rev. Lett. {\bf 120}, 132501 (2018).

\bibitem{PRL-Cupid0}
\textbf{CUPID-0} Collaboration, O.~Azzolini {\em et~al.},
\newblock Phys. Rev. Lett. {\bf 120}, 232502 (2018).

\bibitem{AMoRE}
\textbf{AMoRE} Collaboration, V.~Alenkov {\em et~al.},
\newblock The European Physical Journal C {\bf 79}, 791 (2019).

\bibitem{Echo}
\textbf{ECHo} Collaboration, L.~Gastaldo {\em et~al.},
\newblock The European Physical Journal Special Topics {\bf 226}, 1623 (2017).

\bibitem{Holmes}
\textbf{HOLMES} Collaboration, A.~Giachero {\em et~al.},
\newblock Journal of Instrumentation {\bf 12}, C02046 (2017).

\bibitem{Kittel1974}
C.~Kittel,
\newblock {\em Introduction to Solid State Physics} (Wiley Eastern, New Delhi,
  1974).

\bibitem{20101}
P.~M. Martin,
\newblock {\em Handbook of Deposition Technologies for Films and Coatings},
  Third edition ed. (William Andrew Publishing, Boston, 2010).

\bibitem{AudiaTEc_1}
M.~Schreck {\em et~al.},
\newblock Scientific Reports {\bf 7}, 44462 (2017).

\bibitem{audiatec}
https://www.audiatec.de/,
\newblock Accessed: \today{}.

\bibitem{Rothe2018}
J.~Rothe {\em et~al.},
\newblock Journal of Low Temperature Physics  (2018).

\bibitem{MeVscale}
\textbf{CRESST} Collaboration, G.~Angloher {\em et~al.},
\newblock The European Physical Journal C {\bf 77}, 637 (2017).

\bibitem{ANGLOHER2009}
\textbf{CRESST} Collaboration, G.~Angloher {\em et~al.},
\newblock Astroparticle Physics {\bf 31}, 270  (2009).

\bibitem{OptimumFilter_GM}
E.~Gatti and P.~F. Manfredi,
\newblock La Rivista del Nuovo Cimento (1978-1999) {\bf 9}, 1 (1986).

\bibitem{probst1995model}
F.~Pr{\"o}bst {\em et~al.},
\newblock Journal of low temperature physics {\bf 100}, 69 (1995).

\bibitem{Mancuso:2018zoh}
M.~Mancuso {\em et~al.},
\newblock Nuclear Instruments and Methods in Physics Research Section A {\bf
  940}, 492 (2019), 1711.11459.

\end{thebibliography}

\end{document}